\documentclass[apl,amsmath,amssymb,reprint,superscriptaddress]{revtex4-1}

\usepackage{graphicx}% Include figure files\\
\usepackage[francais,english]{babel}
\usepackage[utf8]{inputenc} 
\usepackage{dcolumn}% Align table columns on decimal point
\usepackage{bm}% bold math
\usepackage{gensymb}
\usepackage[load-configurations = abbreviations]{siunitx}
\usepackage[mathlines]{lineno}% Enable numbering of text and display math
\begin{document}
\sisetup{range-phrase=-}
\preprint{AIP/123-QED}

\title[Sample title]{Nanomechanical resonators based on adiabatic periodicity-breaking in a superlattice}% Force line breaks with \\

\author{F.R. Lamberti}
\thanks{These authors contributed equally to this work}
\affiliation{Centre de Nanosciences et de Nanotechnologies, CNRS, Univ. Paris-Sud, Université Paris-Saclay, C2N – Marcoussis, 91460 Marcoussis, France }%
\affiliation{Laboratoire Matériaux et Phénomènes Quantiques, UMR CNRS 7162, Université Paris-Diderot, 5 Rue Thomas Mann, 75205 Paris, France}%
\author{M. Esmann}
\thanks{These authors contributed equally to this work}
\affiliation{Centre de Nanosciences et de Nanotechnologies, CNRS, Univ. Paris-Sud, Université Paris-Saclay, C2N – Marcoussis, 91460 Marcoussis, France }%
\author{A. Lema\^itre}
\affiliation{Centre de Nanosciences et de Nanotechnologies, CNRS, Univ. Paris-Sud, Université Paris-Saclay, C2N – Marcoussis, 91460 Marcoussis, France }%
\author{C. Gomez Carbonell}
\affiliation{Centre de Nanosciences et de Nanotechnologies, CNRS, Univ. Paris-Sud, Université Paris-Saclay, C2N – Marcoussis, 91460 Marcoussis, France }%
\author{O. Krebs}
\affiliation{Centre de Nanosciences et de Nanotechnologies, CNRS, Univ. Paris-Sud, Université Paris-Saclay, C2N – Marcoussis, 91460 Marcoussis, France }%
\author{I. Favero}
\affiliation{Laboratoire Matériaux et Phénomènes Quantiques, UMR CNRS 7162, Université Paris-Diderot, 5 Rue Thomas Mann, 75205 Paris, France}%
\author{B. Jusserand}
\affiliation{Institut des Nanosciences de Paris, UMR CNRS 7588, Université Pierre et Marie Curie, 4 Place Jussieu 75252 Paris, France}%
\author{P. Senellart}
\affiliation{Centre de Nanosciences et de Nanotechnologies, CNRS, Univ. Paris-Sud, Université Paris-Saclay, C2N – Marcoussis, 91460 Marcoussis, France }%
\author{L. Lanco}
\affiliation{Centre de Nanosciences et de Nanotechnologies, CNRS, Univ. Paris-Sud, Université Paris-Saclay, C2N – Marcoussis, 91460 Marcoussis, France }%
\affiliation{Université Paris Diderot – Paris 7, 75205 Paris CEDEX 13, France }%
\author{N.D. Lanzillotti-Kimura}
\email{daniel.kimura@c2n.upsaclay.fr}
\affiliation{Centre de Nanosciences et de Nanotechnologies, CNRS, Univ. Paris-Sud, Université Paris-Saclay, C2N – Marcoussis, 91460 Marcoussis, France }%

\date{\today}% It is always \today, today,

\begin{abstract}
We propose a novel acoustic cavity design where we confine a mechanical mode by adiabatically changing the acoustic properties of a GaAs/AlAs superlattice. By means of high resolution Raman scattering measurements, we experimentally demonstrate the presence of a confined acoustic mode at a resonance frequency around 350 GHz.  We observe an excellent agreement between the experimental data and numerical simulations based on a photoelastic model. We demonstrate that the spatial profile of the confined mode can be tuned by changing the magnitude of the adiabatic deformation, leading to strong variations of its mechanical quality factor and Raman scattering cross section.  The reported alternative confinement method could lead to the development of a novel generation of nanophononic and optomechanical systems.
\end{abstract}

%Use showkeys class option if keyword
\maketitle

Acoustic cavities confine mechanical vibrations in one or more directions of space \cite{refId0}. They have many applications in the development of novel devices able to generate, manipulate and detect high frequency acoustic phonons \cite{doi:10.1063/1.2178415, PhysRevLett.96.215504}. Furthermore, such systems are at the core of the development of novel optomechanical devices \cite{PhysRevLett.118.263901}.  Well established designs of one-dimensional acoustic nanocavities are phononic Fabry Perot resonators capable of operating in the technologically relevant sub-THz range \cite{PhysRevLett.89.227402, LANZILLOTTIKIMURA201580, PhysRevB.84.115453}. They are built out of highly reflective acoustic distributed Bragg reflectors (DBR) \cite{Jusserand1989, Narayanaamurti717} obtained  by stacking materials with different elastic properties in a periodic way. Most of the mechanical properties of an acoustic DBR can be described by an acoustic band diagram \cite{PhysRevB.31.2080}: acoustic minigaps are opened at the center and at the edge of the superlattice Brillouin zone, and the acoustic modes can be described in the Bloch mode formalism \cite{PhysRevB.35.2808}. By introducing a defect such as a spacer inside an acoustic DBR,  acoustic Fabry-Perot cavities are obtained  \cite{PhysRevLett.89.227402}. Acoustic confinement can be probed by performing Raman scattering spectroscopy and pump probe experiments \cite{PhysRevLett.89.227402, PhysRevLett.110.037403, PhysRevLett.98.265501, PhysRevLett.102.015502}. Moreover, stacking   several of these resonators has opened up the possibility  to engineer and study the dynamics of complex phononic systems \cite{PhysRevLett.104.197402, PhysRevB.75.024301}.

So far,  sub-THz phononics has extensively used the standard Fabry-Perot approach and very little work has been dedicated to the development of other designs \cite{PhysRevB.92.020404, PhysRevB.92.165308, PhysRevLett.104.187402}. This is in great contrast with their optical counterparts, for which sophisticated optical cavity designs have emerged over the years, providing stronger spatial confinement, higher quality factors as well as reduced sensitivity to nano fabrication imperfections \cite{akahane_high-q_2003, doi:10.1063/1.1354666, test}. One elegant design proposed in the optical domain consists in introducing tapered regions where the photonic crystal periodicity is adiabatically broken. Such an approach allows for reduced  optical losses---hence increased  optical Q-factors---when going to 3D confinement. This strategy has been adopted in several optical systems,  such as 2-dimensional photonic crystal membranes \cite{song_ultra-high-q_2005}, nanobeams \cite{chan_laser_2011, doi:10.1063/1.3579535} and waveguides \cite{Velha:07} or micropillars \cite{PhysRevLett.108.057402, Zhang:09}. 

In this Letter, we report the design and the experimental measurement of the confinement properties of an adiabatic acoustic cavity, designed to operate at a resonance frequency of $\approx$ 350 GHz. By progressively changing the periodicity of an acoustic DBR, we adiabatically transform the acoustic band diagram of the system, leading to the generation of a confined mechanical state. The presented results were obtained on a sample where the DBR periodicity was adiabatically transformed with a maximum amplitude of approximately 7\%, which is technologically challenging to fabricate, even by molecular beam epitaxy (MBE). We probed the presence of the confined phononic state by performing high resolution Raman scattering experiments. Furthermore, by changing the magnitude of the adiabatic transformation, we demonstrate that we can significatively transform the spatial profile of the confined mode, leading to major changes in its mechanical Q-factor and Raman scattering cross section.
Such kind of design could lead, as in the case of optics, to the development of new 3 dimensional mechanical resonators with quality factors overcoming the ones currently achieved with standard Fabry-Perot designs \cite{PhysRevLett.118.263901,2017arXiv170708687L}.

We start the conception of the adiabatic cavity by designing an acoustic GaAs/AlAs DBR constituted by 29 GaAs/AlAs layer pairs. The layer thicknesses of AlAs and GaAs are 12 nm and 3.4 nm, respectively. By choosing these layer thicknesses, we obtain a ($\frac{\lambda}{4},\frac{3\lambda}{4}$) acoustic DBR, where $\lambda$ corresponds to the wavelength of the acoustic phonons in GaAs and AlAs respectively, for a frequency of \SI{354}{\giga\hertz} \cite{Jusserand1989}. Then, by gradually changing the layers' thicknesses (Fig. 1a, top panel), we introduce an adiabatic perturbation at the center of the structure. The envelope of the perturbation has the shape of a $sin^2$ function, an amplitude of 7\%, and extends over 12.5 layers pairs. We compute the reflectivity of the system around \SI{350}{\giga\hertz} embedded in a GaAs matrix, as shown with the black curve in Fig. 1b. We note the presence of a sharp dip inside the stop band of the system at a frequency of \SI{353}{\giga\hertz}, corresponding to a confined mode. The spatial profile of the acoustic mode in the adiabatic cavity is shown in Fig. 1a (bottom panel). It is determined by solving the equation for one dimensional propagation of longitudinal acoustic waves using transfer matrix calculations. It is confined at the center of the structure and decays exponentially when we move away from the adiabatically perturbed region. The red dashed line in Fig. 1b represents the simulated reflectivity spectrum of the considered DBR without any adiabatic perturbation. The high reflectivity region corresponds to the first minigap at the zone center of the Brillouin zone.

The presence of a confined state can be explained by locally applying the Bloch mode formalism in the aperiodic part of the sample, and in particular for one period of alternating AlAs/GaAs layer pairs \cite{0268-1242-16-3-201, PhysRevLett.108.057402, PhysRevB.75.024301}. We calculate for every pair of AlAs/GaAs layers the corresponding local acoustic band diagram. In the inset of Fig. 1a (bottom panel), we show the position of the first zone center band gap as function of the position in the sample. The eigenfrequency of the confined mode is represented by the horizontal dashed line. By progressively increasing the size of the layers, we gradually redshift the position of the local acoustic bandgap of the system. At the center of the perturbed region, the confined mode is outside the bandgap and is therefore allowed to propagate. However, by moving away from the center, the mode enters adiabatically into the bandgap and is progressively reflected by the DBRs, leading to its confinement.

\begin{figure}[!htbp]
\includegraphics[width=0.5\textwidth]{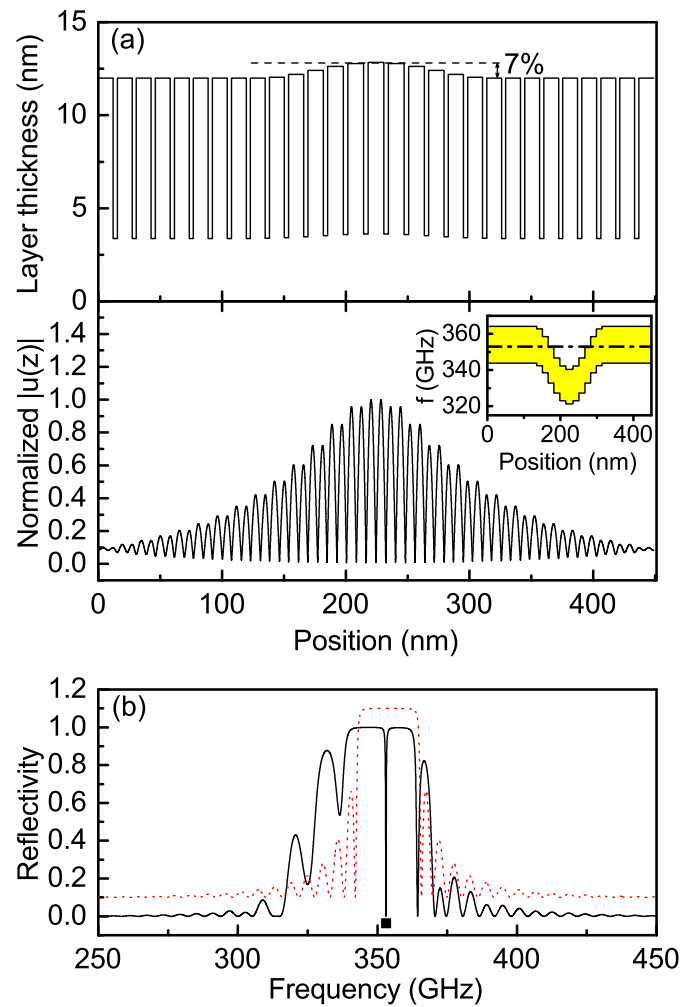}% Here is how to import EPS art
\caption{\label{fig:epsart} (Color online)
\textbf{(a) top panel}: Layer thickness evolution as function of the position in the sample.
\textbf{(a) bottom panel}: Spatial profile of the confined acoustic mode inside the structure (eigenfrequency: \SI{353}{\giga\hertz}).Inset: evolution of the local acoustic minigap as function of the position in the sample.
\textbf{(b)}: Acoustic reflectivity for the adiabatic nanocavity around \SI{350}{\giga\hertz}. The spectral position of the confined mode plotted in (a) (bottom panel)  is marked by a black square.}
\end{figure}

A GaAs/AlAs-based sample was fabricated by molecular beam epitaxy on a (001) GaAs substrate. The adiabatic acoustic cavity was characterized by Raman scattering spectroscopy performed at room temperature. We used a Ti-sapphire tunable laser set at a wavelength of \SI{913}{\nm}, and the collected spectra were dispersed using a double HIIRD2 Jobin Yvon spectrometer equipped with a liquid $N_2$ charged coupled device (CCD). The adiabatic acoustic cavity is embedded between two optical $\text{Al}_{0.1}\text{Ga}_{0.9}\text{As}/\text{Al}_{0.95}\text{Ga}_{0.05}\text{As}$ DBRs and constitutes the $\frac{3\lambda_0}{2}$ spacer of an optical microcavity ($\lambda_0$ corresponds to the wavelength of the confined optical mode).  The top (bottom) optical DBR is constituted by 14 (18) pairs. Fabricating the acoustic structure inside an optical cavity not only enhances the Raman scattering signals up to a factor $10^5$ as shown in \cite{PhysRevLett.75.3764, PhysRevLett.86.3411, PhysRevB.53.R13287, PhysRevB.57.2402}, but it also modifies the Raman scattering selection rules, allowing to detect Raman signals associated to the confined mode in a backscattering experimental configuration. We collected the scattered light at normal incidence, whereas the excitation laser was incident with and angle different from  \SI{90}{\degree}. As the sample has been grown with a thickness gradient, we could tune the resonance frequency of the collection mode between $\approx$ \SIrange[range-units=single]{0.8}{1.0}{\micro\metre} by changing the position of the laser spot. When the frequency of the scattered photons corresponded to the energy of the collection mode, we were in a condition of single optical resonance. We further enhanced the intensity of Raman signals by taking advantage of the in-plane dispersion relation of the optical cavity: by carefully changing the incidence angle of the laser, it was possible to set the incoming photons in resonance with the excitation mode. When both the spot position and the laser incidence angle were set in order to maximize a Raman signal, we were in condition of double optical resonance (DOR) \cite{PhysRevLett.75.3764,PhysRevB.79.035404}.

In Fig 2.a (Black curve) we show the simulated acoustic band diagram of the  ($\frac{\lambda}{4},\frac{3\lambda}{4}$) GaAs/AlAs acoustic DBR used for the conception of our sample, without any adiabatic defect. The grey area highlights the spectral interval of the first zone-centre acoustic minigap. The vertical orange solid line indicates the spectral position of the dip in reflectivity marked by a black square in Fig. 1a (bottom panel), corresponding to the resonance frequency of the confined mechanical mode.

The measured Raman spectrum is presented in Fig. 2b. The frequency of the inelastically scattered light is $f_{laser}+\Delta f$, where $\Delta f$ is the frequency shift introduced during the Raman process and $f_{laser}$ is the frequency of the incident laser. The DOR condition was optimized for maximizing Raman signals for $\Delta f$ $\approx$ \SI{350}{\giga\hertz}. We observe 4 clear Raman peaks in the measured spectrum. The most intense one is well located in the frequency interval of the zone-centre acoustic minigap, also marked in Fig. 2b by a grey area. This Raman peak is generated by the cavity mode (CM) confined in the adiabatic structure.

\begin{figure}[htbp]
\includegraphics[width=0.5\textwidth]{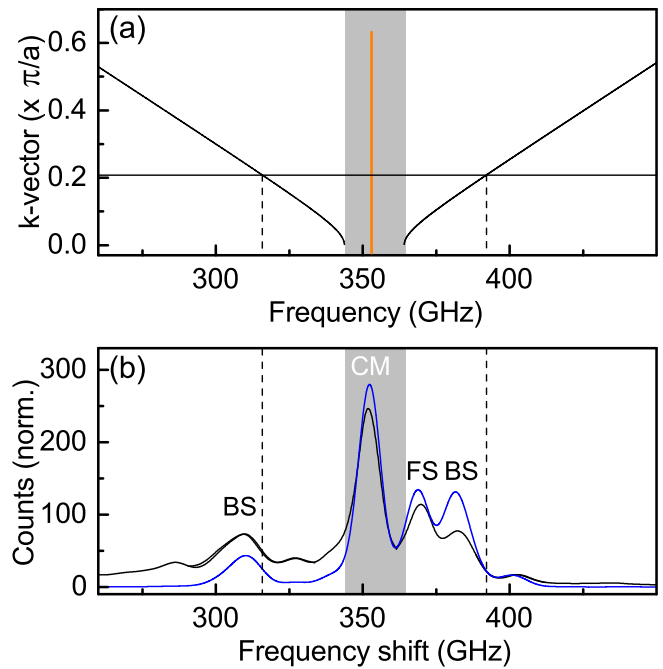}% Here is how to import EPS art
\caption{\label{fig:epsart}(Color online)
\textbf{(a)}: Calculated acoustic band diangram for the DBR without any adiabatic defect (black curve). The frequency interval corresponding to the zone centre acoustic minigap is marked by a grey area. The orange solid line corresponds to the resonance frequency of the mechanical confined mode in the adiabatic cavity.
\textbf{(b)}: Measured Raman spectrum (black curve), simulated Raman spectrum (blue curve) and spectral position of the zone centre acoustic minigap (grey area). For the measured Raman spectrum, the peak corresponding to the confined mode is marked by the label CM.}

\end{figure}

We implemented a photoelastic model to calculate the Raman spectrum of the structure \cite{PhysRevB.37.4086, RUELLO201521,PhysRevB.75.245303, PhysRevLett.98.265501,PhysRevLett.98.265501, Jusserand1989}. The simulated Raman spectrum is shown in Fig 2b. (blue curve), after convoluting it with a gaussian curve to account for the experimental resolution (\SI{3.5}{\giga\hertz}). We note that the simulated spectrum perfectly reproduces all the features of the experimental data, accounting for the good quality of the sample growth process \cite{PhysRevB.66.125311}. The peaks located around \SI{310}{\giga\hertz} and \SI{380}{\giga\hertz} would be normally observable in a back scattering (BS) geometry in structures with no optical confinement. These mechanical modes are localized in the DBRs (propagative modes). The mode at \SI{369}{\giga\hertz}, on the contrary, is usually active in forward scattering (FS) geometry in the absence of optical confinement. Small differences between the experiments and simulations in the relative intensities can be attributed to optical resonant effects.

The vertical dashed lines in Fig. 2a and Fig. 2b correspond to the condition $q = 2k_{laser}$, where $k_{laser}$ corresponds to the wavenumber of the incident laser. They indicate the frequencies of mechanical modes that are usually Raman active in a BS geometry for a superlattice \cite{PhysRevB.37.4086,PhysRevB.31.2080}. We observe that in the measured Raman spectrum, the peaks associated to back scattering are red shifted with respect to these frequencies. The introduction of an adiabatic defect in a superlattice also affects the spectral position of the Raman peaks associated to propagative modes, as it increases the average thicknesses of the layers at the centre of the structure.

We numerically investigated the confinement properties of the adiabatic cavity by exploring the impact of the adiabatic $sin^2$ transformation inside the structure. We define $\alpha$ as the maximal adiabatic transformation introduced in the system. For the case of the experimentally studied cavity in this letter,  $\alpha = 7\%$. In Fig. 3a, the black curve (bottom curve) corresponds to the simulated reflectivity curve of the adiabatic cavity around \SI{350}{\giga\hertz} for $\alpha = 7\%$. We then progressively increase the magnitude of $\alpha$. We observe that the dip corresponding to the confined mode is gradually red-shifted inside the acoustic stop band. By further increasing $\alpha$ a second sharp dip in the acoustic stop band appears, evidencing the presence of a second confined mode. Both dips are clearly visible for $\alpha = 11\%$ as shown with the green reflectivity curve (middle curve) in Fig. 3a, for which we have introduced a vertical offset for clarity. Eventually, by raising $\alpha$ up to 15\%, the first mode disappears in the Bragg oscillations of the system, and the second mode reaches the centre of the acoustic stop band (top red curve in Fig. 3a).

\begin{figure}[htbp!]
\includegraphics[width=0.5\textwidth]{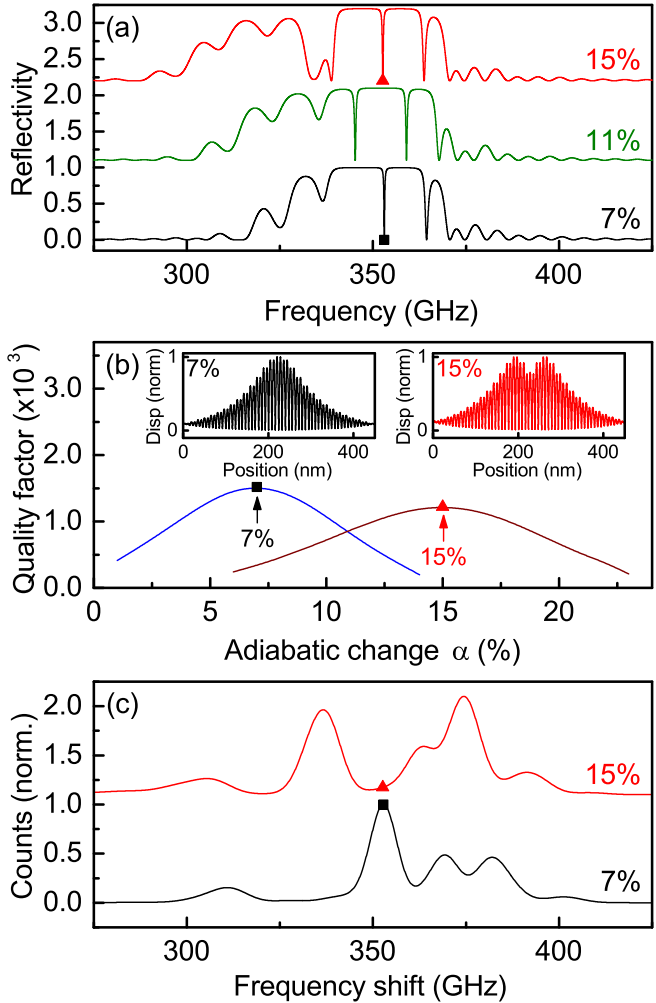}% Here is how to import EPS art
\caption{\label{fig:epsart}(Color online)
\textbf{(a)}: Calculated reflectivity curves around 350 GHz for different adiabatic transformations ($\alpha= 7 \%$, black curve, $\alpha = 11\%$, green curve, $\alpha = 15\%$, red curve). Offsets between the curves have been introduced for clarity.
\textbf{(b)}: Evolution of the mechanical quality factor of the fundamental confined mode (black curve) and first harmonic (brown curve) as function of the adiabatic transformation. The points corresponding to $\alpha= 7 \%$ and $\alpha = 15\%$ are marked by a black square and red triangle, respectively. Insets: spatial profile of the fundamental confined mode (black curve) and first harmonic mode (red curve) calculated for 7\% and 15\% adiabatic changes respectively.
\textbf{(c)}: Simulated Raman scattering spectra for adiabatic transformations of 7\% (black curve) and 15\% (red curve, with offset). The symbols represent the spectral positions of the phononic confined mode inside the system.}
\end{figure}

The spatial profiles of the first and second confined modes are plotted in the insets of Fig. 3b, calculated respectively for $\alpha = 7\%$ and $\alpha = 15\%$. Both modes are confined at the centre of the structure. However, the second mode presents two maxima in its displacement pattern. The first mode corresponds to the fundamental confined mode (already plotted in Fig. 1a bottom panel), and the second to the first harmonic of this acoustic equivalent of a quantum well. It is therefore possible to select the desired spatial profile of the mode which is optimally confined by changing the amplitude of the adiabatic deformation, and to finely tune its mechanical resonance frequency.

To characterize the resonator mechanical performance, we studied the evolution of the confinement properties of the two considered modes as function of the amplitude of adiabatic transformation $\alpha $ (Fig. 3b)). The values of the mechanical quality factors (Q-factors) increase when the resonance frequencies approach the centre of the acoustic minigap.  Maximal values for the mechanical Q-factors for the fundamental and first harmonic modes are reached for 7\% ($Q_{mechanical}=1520$) and 15\% ($Q_{mechanical}=1220$) adiabatic transformations respectively, marked by a black square and a red triangle respectively in Fig. 3b.

To compare this design to a standard Fabry-Perot cavity, we have simulated the Q-factor of an acoustic Fabry-Perot resonator composed of 14 $\frac{\lambda}{4},\frac{3\lambda}{4}$ GaAs/AlAs layer pairs for each DBR, and one $\frac{\lambda}{2}$ AlAs spacer. This structure contains the same number of layers as for the adiabatic system. The Q-factor reached is 1570, very close to the value of the Q-factor reached for $\alpha = 7\%$.
 
In Fig. 3c we plot the simulated Raman spectra around \SI{350}{\giga\hertz} for cavities with $\alpha = 7\%$ (black curve) and $\alpha = 15\%$ (red curve with offset). We have marked by a black square the Raman peak corresponding to the presence of the first confined mode. As it has been shown in Fig. 2, the confined phonons in the adiabatic structure are Raman active. For $\alpha = 15\%$ (red curve), we marked the resonance frequency of the second harmonic confined mode with a red triangle. The confined mode induced by a $\alpha = 15\%$ perturbation presents a different symmetry in strain, resulting in a Raman inactive mode, as indicated in Fig. 3c with a triangle. By tuning the paramater $\alpha$ it is thus possible to tailor the spatial profile of the adiabatic confined mode and its symmetry.

In conclusion, we demonstrated the adiabatic confinement of longitudinal acoustic phonons at a resonance frequency of \SI{350}{\giga\hertz} by progressively breaking the periodicity of an acoustic superlattice. We probed the presence of a confined mode by performing Raman scattering spectroscopy experiments in a DOR configuration. Numerical simulations based on transfer matrix calculations and a photoelastic model well reproduce our experimental data, accounting for the high quality of the MBE grown sample and showing the feasibility of actually fabricating these systems. We investigated the impact of the adiabatic transformation magnitude on the spatial profile of the confined modes and on their mechanical quality factors. The presented adiabatic cavity is one of the first steps in the study of acoustic phonon resonators where a local strain engineering is performed. As it has already been demonstrated for standard acoustic Fabry-Perot designs \cite{2017arXiv170708687L,PhysRevLett.118.263901}, it is possible to fabricate out of these planar structures 3 dimensional optomechanical microresonators operating at extremely high mechanical frequencies, and for which the confined mechanical and optical modes strongly interact. Combining the simultaneous localization of photons and phonons, the reported system has the potential of being at the heart of a novel generation of optomechanical resonators based on DBR structures.

\begin{acknowledgments}
This work was partially supported by a public grant overseen by the French National Research Agency
(ANR) as part of the ”Investissements d’Avenir” program (LabexNanoSaclay, reference: ANR-10-
LABX- 0035), the ERC Starting Grant No. 715939 NanoPhennec, the French Agence Nationale pour la Recherche (grant ANR QDOM), the French RENATECH network.
\end{acknowledgments}%

%\nocite{*}
\bibliography{New_Biblio_Article_Adiab}% Produces the bibliography via BibTeX.

\end{document}